\documentclass[prd,twocolumn,amsmath,amssymb,floatfix,superscriptaddress,nofootinbib]{revtex4-1}

\usepackage{graphicx}
\usepackage{amssymb}
\usepackage{amsmath}
\usepackage{bm}
\usepackage{color}

\DeclareMathAlphabet\mathbfcal{OMS}{cmsy}{b}{n}
\definecolor{darkgreen}{RGB}{50,150,0}
\definecolor{orange}{RGB}{255,127,80}
\definecolor{purple}{cmyk}{0.5,1.0,0,0}

\def\be{\begin{equation}}
\def\ee{\end{equation}}
\def\ben{\begin{equation} \nonumber}
\def\een{\end{equation}}
\def\ban{\begin{eqnarray*}}
\def\ean{\end{eqnarray*}}
\def\ba{\begin{eqnarray}}
\def\ea{\end{eqnarray}}
\def\({\left(}
\def\){\right)}
\newcommand{\bs}[1]{\boldsymbol{#1}}

\newcommand{\Comment}[1]{{}}

\definecolor{ultramarine}{rgb}{0.07, 0.04, 0.56}
\definecolor{cadmiumgreen}{rgb}{0.0, 0.42, 0.24}
\definecolor{indigo(dye)}{rgb}{0.0, 0.25, 0.42}
\usepackage[linktocpage=true]{hyperref}
\hypersetup{
colorlinks=true,
citecolor=ultramarine,
linkcolor=cadmiumgreen,
urlcolor=indigo(dye),
pdfauthor={},
pdftitle={},
pdfsubject={}
}

\begin{document}

\title{Can transition radiation explain the ANITA event 3985267?}
\author{Pavel Motloch}
\affiliation{Kavli Institute for Cosmological Physics \& Department of Physics, University of Chicago, Chicago, IL
60637, USA}
\author{Jaime Alvarez-Mu\~niz}
\affiliation{Departamento de F\'isica de Part\'iculas \& Instituto Galego de F\'isica de
Altas Enerx\'ias, Universidade de Santiago de Compostela, 15782 Santiago de Compostela,
Spain}
\author{Paolo Privitera}
\affiliation{Kavli Institute for Cosmological Physics \& Department of Physics, University of Chicago, Chicago, IL
60637, USA}
\author{Enrique Zas}
\affiliation{Departamento de F\'isica de Part\'iculas \& Instituto Galego de F\'isica de
Altas Enerx\'ias, Universidade de Santiago de Compostela, 15782 Santiago de Compostela,
Spain}

\begin{abstract}
\noindent
We investigate whether transition radiation from a particle shower crossing 
the interface between Earth and air and induced by an Earth-skimming
neutrino can explain the upward event announced recently by the ANITA Collaboration.
While the properties of the observed signal can in principle be explained with
transition radiation, a conservative upper limit on the experiment's aperture
for this kind of signal shows that the flux necessary for a successful explanation is in tension with the
current best limits from the Pierre Auger Observatory, the IceCube neutrino detector and the 
ANITA balloon.
We also show that in this scenario, the direction of the incoming neutrino is
determined precisely to within a few degrees, combining the polarization properties of the observed events
with the Earth opacity to ultra high energy neutrinos. 
\end{abstract}

\maketitle

\section{Introduction}

Ultra high energy cosmic rays (UHE CR) have been detected with energies 
above $10^{20}$ eV \cite{Abraham:2010mj,Fukushima:2015bza}. They can interact with
photons of the cosmic microwave background, what limits the distance to which we can in
principle detect UHE CR sources (see e.g. \cite{Watson:2013cla} for a review). As a
byproduct of these
interactions, cosmogenic neutrinos with energies of several EeV are created.

Many ongoing and proposed experiments, such as IceCube \cite{Aartsen:2015zva}, ARA \cite{Allison:2011wk}, ARIANNA
\cite{Barwick:2014pca} and the Pierre Auger Observatory \cite{Aab:2015kma}, have
capabilities to search for these cosmogenic neutrinos. To date, we are
still awaiting a first detection. An original approach is pursued by the ANITA
collaboration, who operate a suborbital balloon flying above Antarctica
\cite{Gorham:2008dv}. During each flight which lasts about a month, the polarization antennas
mounted on this balloon observe a large portion of the Earth surface, searching for transient
signals consistent with those predicted from neutrino-induced particle showers. From the
data collected during the first two flights, ANITA collaboration sets the tightest constraints on the
neutrino flux at energies above about 40 EeV \cite{Gorham:2010kv, Aartsen:2015zva}.

As pointed out in \cite{Hoover:2010qt}, the ANITA balloon can also serve as an UHECR detector
by detecting the radio signal emitted by the UHECR atmospheric showers. During the first
flight, 16 such events were detected with polarization consistent with a geomagnetic
origin of the signal. Among these events, 14 were seen after reflection off the ice
surface of the continent while 2 events were seen directly, without reflection on
the surface. In the second flight of the balloon, a new ``direct event'' was
observed \cite{Gorham:2016zah}.

In a recent reanalysis of the data \cite{Gorham:2016zah}, the ANITA collaboration found an
additional event (event 3985267) which shares some properties of the signals emitted by the atmospheric
showers. The detected electric field was predominantly horizontally polarized, with the polarization direction
consistent with the geomagnetic origin of the emission \cite{Huege:2016veh}. Its waveform does not correlate well
with the reflected CR signal, which is why it was neglected in the previous analyses. 
The waveform of the event can be interpreted as coming from an atmospheric shower observed directly
because it lacks the change of sign in the polarization expected for a signal 
from a cosmic-ray induced shower reflected in the ice at a large zenith angle.
However, this interpretation faces difficulties as the signal arrives from a direction which is
$27.4^\circ$ below the horizontal. The geomagnetic radio emission from an air shower is known to be 
beamed roughly in the
direction of the incoming primary particle \cite{Huege:2016veh} and the shower would then have to be initiated by a
particle which has crossed through a significant portion of the Earth. Neutrinos are the only known
particles capable of doing this. The authors of \cite{Gorham:2016zah} considered the
case of a $\tau$ neutrino producing a $\tau$ lepton which then decays in the atmosphere 
and initiates an atmospheric shower, but found that the probability of this occurring is 
suppressed by a factor $\sim 4 \cdot 10^{-6}$ due to neutrino absorption 
in the Earth induced by the large neutrino interaction cross section at EeV energies.
Anthropogenic origin for the event was found to be rather strongly disfavored by the data.

Recently, we investigated transition radiation (TR) from particle showers crossing the interface
between dense media and air 
\cite{Motloch:2015wca}. It was found that the signal is typically emitted into a wide solid angle and is
coherent up to frequencies of several GHz. 
Because of this broad emission, the signal seen at $27.4^\circ$
below the horizon could be explained as transition radiation from an Earth-skimming neutrino, 
avoiding the large suppression due to 
neutrino absorption in the Earth at EeV energies. 
In this paper we investigate whether
it is realistic that the upcoming ANITA event was caused by transition radiation
emitted by a neutrino-induced shower which starts its development in ice and then crosses
into the atmosphere.  While the transition radiation signal is about an order of magnitude
weaker than the Askaryan emission, the additional solid angle might
explain why the first detection of a $\sim$ EeV neutrino could occur through 
transition radiation. 

In the first part of the paper we discuss whether it is possible to explain
the parameters of the observed ANITA event in terms of transition radiation. In the second
part we then place a conservative upper limit on the exposure of the ANITA balloon to this kind of events
and compare it with exposures of other experiments. We conclude by discussing our findings.

\section{Can Transition Radiation Explain The Main Features Of The Event?}
\label{sec:TRExplanationPossible}

In this section we evaluate whether it is in principle possible to explain the observed 
upward-going event
in terms of TR.
We work with the approximation of a spherical Earth of radius $R =
6373\,\mathrm{km}$ and the height of the balloon above the surface $h =
34\,\mathrm{km}$. Given the order-of-estimate nature of our arguments we will not discuss
uncertainties.

\subsection{Geometry}

\begin{figure}
\center
\includegraphics[width = 0.48 \textwidth]{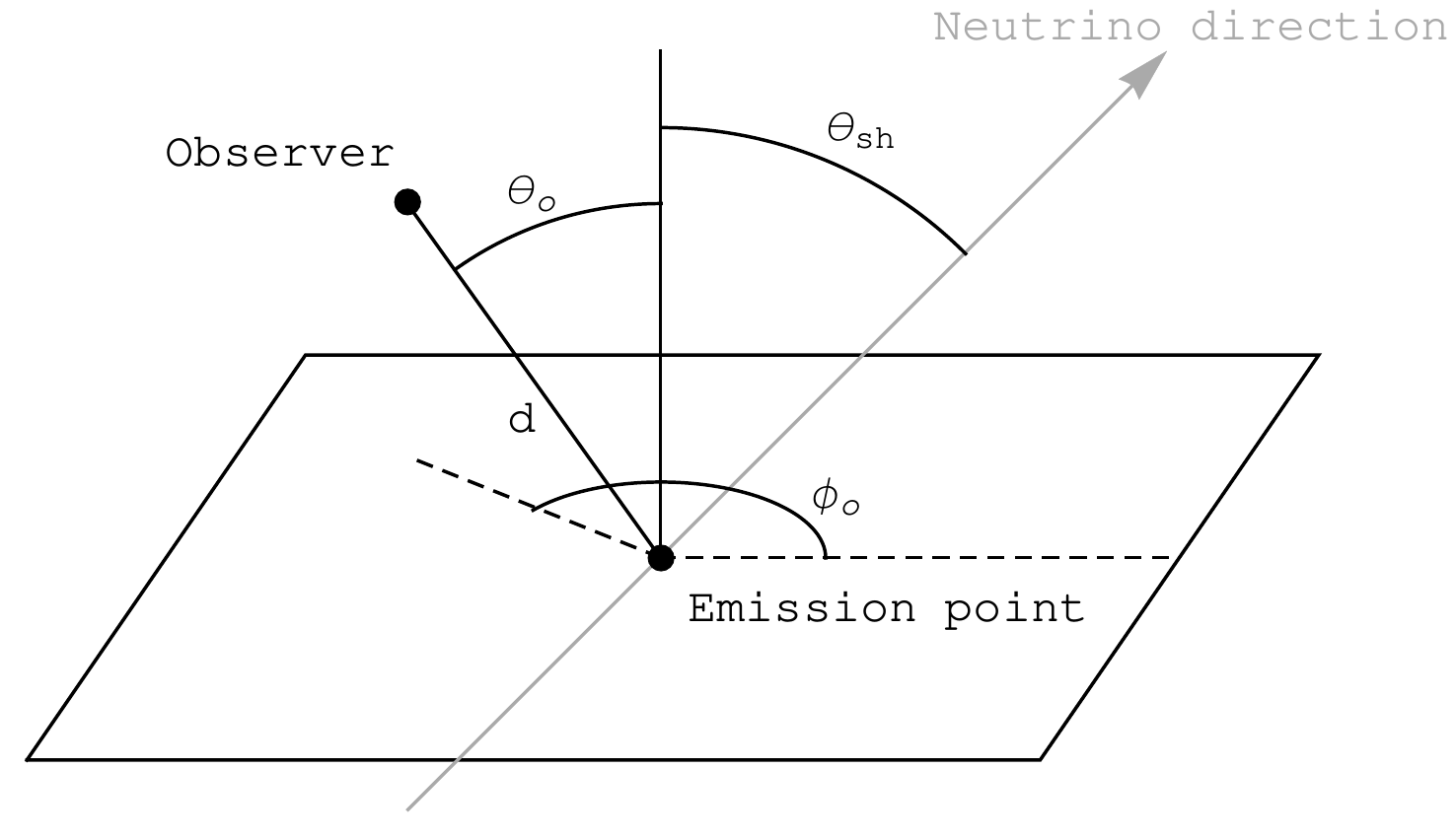}
\caption{Sketch of the emission geometry. The neutrino moves along a trajectory (gray
arrow) described by a zenith angle $\theta_\mathrm{sh}$ from the vertical; the emission point is
at the intersect of this trajectory with the Earth's surface. The observer is located at distance $d$
from the emission point at zenith angle $\theta_o$. Projections of neutrino and observer
directions into the horizontal plane are displayed with dashed lines; the angle between
them is $\phi_o$.
}
\label{fig:sketch}
\end{figure}

Assuming a transition radiation origin of the signal, we have to locate its emission on the
Earth's surface. Given the distance to the balloon, we will approximate the emission as
coming from a single point, neglecting its spatial extent. 
For the ANITA upcoming event, the observer's zenith angle $\theta_o$ 
shown in the sketch in Fig.~\ref{fig:sketch} can be calculated from simple trigonometry 
giving $\theta_o = 63.2^\circ$.
This corresponds to a distance $d$ between the balloon and the emission point (see Fig.~\ref{fig:sketch})
of about $d \approx 75\, \mathrm{km}$.

\begin{figure}
\center
\includegraphics[width = 0.48 \textwidth]{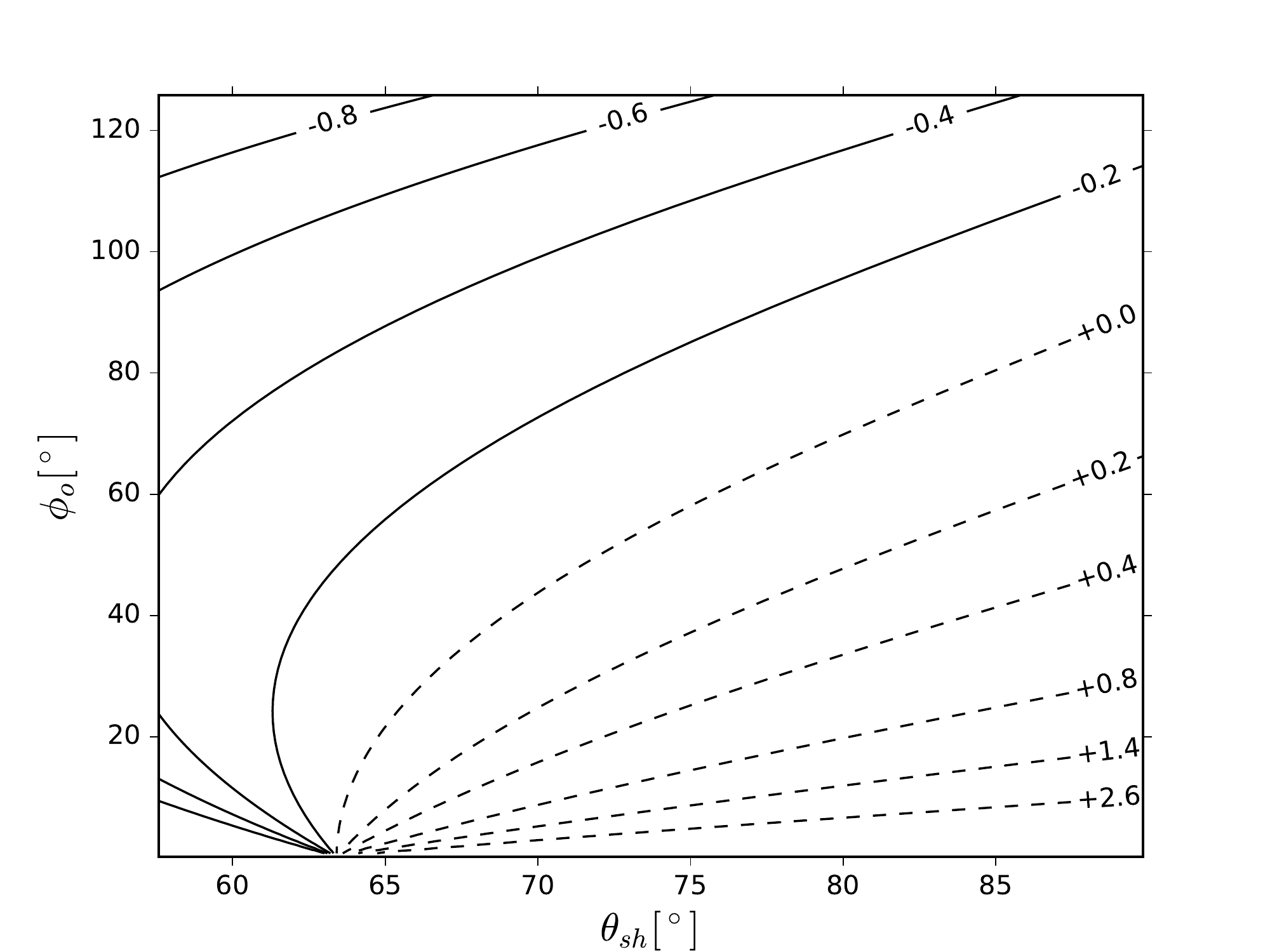}
\caption{Approximate degree of polarization of the signal $A_V/A_H$ defined in Eq.~(\ref{eq:deg_pol}) 
as a function of the
neutrino inclination $\theta_\mathrm{sh}$ and the observer position
$\phi_o$ (see Fig~\ref{fig:sketch}); using fixed $\theta_o = 63.2^\circ$ determined by the
data. Only portion of the parameter space in agreement with the measured values $A_V/A_H
\approx 0.2$ is shown.
}
\label{fig:polarization}
\end{figure}

The remaining two angles describing the geometry --- the neutrino zenith angle $\theta_\mathrm{sh}$
and the azimuth $\phi_o$ between the observer and the neutrino directions
(Fig.~\ref{fig:sketch}) --- 
can be constrained using the polarization of the signal. It is a sensible approximation that
most of the shower particles are moving in the direction of the incoming neutrino 
\be
	\bs{\hat{v}} \sim \(\sin \theta_\mathrm{sh}, 0, \cos \theta_\mathrm{sh}\) .
\ee
Emission in the observer direction
\be
	\bs{\hat{k}} = \(\sin \theta_o \cos \phi_o, \sin
	\theta_o \sin \phi_o, \cos \theta_o\)
\ee
is in such case polarized as \cite{Motloch:2015wca}
\be
\label{vperp}
	\bs{E} \sim \bs{v_\perp} = \bs{\hat k} \times \(\bs{\hat{v}} \times \bs{\hat k} \) .
\ee
The electric field $\bs{E}$ is projected onto two orthogonal directions
given by vectors
\ba
	\bs{e_V} &=& \(\cos \theta_o \cos \phi_o, \cos \theta_o
	\sin \phi_o, - \sin \theta_o\)
\ea
and
\ba
	\bs{e_H} &=& \(\sin \phi_o, -\cos \phi_o, 0\)  ,
\ea
the latter in a horizontal direction. 
We can then quantify the ratio of the two polarizations expected for the signal defined here as
\be
	\frac{A_V}{A_H} = \frac{\bs{e_V} \cdot \bs{v_\perp}}{\bs{e_H} \cdot \bs{v_\perp}}.
\label{eq:deg_pol}
\ee
In Fig.~\ref{fig:polarization} we show the 
polarization ratio as a function of $\theta_\mathrm{sh}$ and $\phi_o$ for the known
value of $\theta_o$. 

From Figure 1 of paper \cite{Gorham:2016zah}, the ratio of strengths of the vertical and
horizontal polarizations is approximately $A_V / A_H \approx 0.2$, using the relative
magnitudes of
the components of the electric field at the time of its maximal magnitude. 
As shown in Figure~\ref{fig:polarization}
there are combinations of $(\theta_\mathrm{sh},~\phi_o)$ for which the polarization fraction 
predicted by the TR model matches the experimental data. The observed degree of polarization 
can be explained if $\theta_\mathrm{sh}$ is large, typically $> 65^\circ$. 
This agrees with the interpretation of the ANITA detected event as being Earth-skimming. 
Also as both senses of polarity are possible, TR does not suffer from the sign mismatch which 
disfavors the explanation of the detected event as a reflected UHECR signal.

\subsection{Energy}

We can use the knowledge of the observed peak electric field $E_\mathrm{max} \sim
5 \cdot 10^{-4}\, \mathrm{V/m}$ to put {an approximate} lower bound on the neutrino
energy. ANITA detects signals between 200 MHz and 1200 MHz \cite{Hoover:2010qt},
corresponding to a bandwidth of $\Delta f = 1000\,
\mathrm{MHz}$.
Assuming a mostly flat spectrum in the frequency domain with amplitude $E_\omega$, the
peak value of the observed signal would correspond to 
\be
	E_\omega \approx 5 \cdot 10^{-7}\, \mathrm{V/m/MHz} 
\ee
in the convention of \cite{Motloch:2015wca}.
The signal strength decreases proportionally with the distance between
emission and observation. For the determined observer - emission distance $d\sim 75$ km, this
corresponds to a distance-independent measure of the emission strength
\be
	E_\omega d \approx 37.5\, \mathrm{mV/MHz} .
\ee

In the following we will consider the cases of purely hadronic showers induced
by all flavors of neutrinos in neutral current (NC) interactions, and by tau and muon
neutrinos in charged current (CC) interactions 
(neglecting the showers induced by the tau and muon lepton - see below), as well as 
electromagnetic$+$hadronic showers induced by electron neutrinos in CC interactions.
In \cite{Motloch:2015wca} it was shown that the amplitude of the electric field
due to TR approximately scales with the shower energy for energies in the range of 1 to 100 TeV, but at
1 EeV this scaling was broken because of the Landau-Pomeranchuk-Migdal effect (LPM)
\cite{Landau:1953um}. This effect dramatically reduces the probability of interaction of
electrons and positrons at the highest energies, leading to elongation of
electromagnetic (sub-)showers. 
Due to the elongation, the number of particles present at the
shower maximum is reduced and the TR signal is diminished proportionally. Simulations
show that the integral of the number of electrons over shower length scales very
accurately with energy \cite{Zas:1991jv}; this is why the number of particles at shower maximum
typically scales with the shower energy when there is no LPM effect. When the LPM effect
takes place, an increase in the shower length must be accompanied by a corresponding 
decrease in the average number of particles at shower maximum. 
In the following we adopt a simple model in which the product of the
shower length and the number of particles at shower maximum scales linearly with shower energy.
The TR emission is proportional to the electron excess~\cite{Motloch:2015wca}, which is 
approximately a constant fraction (of order 25\% in ice) of the total number
of electrons and positrons\footnote{The maximal achievable amplitude of the TR
  signal is thus proportional
to the maximum number of particles at shower maximum \cite{Motloch:2015wca}.}. 

For electromagnetic showers at EeV energies, we will model the shower 
length, defined as the length of the shower over which the number of particles remains
above half the value at maximum, as 
\be 
\label{showerLength}
	L_\mathrm{sh,em} = 25 \left(\frac{E_\mathrm{sh}}{\mathrm{EeV}}\right)^{1/3}
	\,\mathrm{m} .
\ee 
This parametrization is taken from \cite{Tartare:2012zz},
  where both the LPM effect and the
  photonuclear cross section are accounted for showers simulated in rock.
The given parameterization has been corrected for ice. 
The radiation length $X_0$ in ice is a factor of four 
larger~\cite{Olive:2016xmw} and the energy above which the showers start
to increase in length increases also by a factor of 4. 
These effects go in the opposite directions and overall one expects showers in ice
to be $\sim 2$ times longer than in rock. However, in
addition in  
Ref.~\cite{Tartare:2012zz} the definition of shower length requires the
number of electrons to exceed only $1/10$ of the maximum. 
A reduction in shower length of about a factor of two can be expected from
changing the definition of shower length from $1/10$ to $1/2$ \cite{AlvarezMuniz:1997sh},
which should better describe our situation and bring the shower length back to
\eqref{showerLength}. 
In any case, the main conclusion of this work does not change if we increase or decrease
the above factor by two or if we change the exponent to $1/2$ as
suggested by an analytic model \cite{Gerhardt:2010bj} using a different
definition of shower length. For our purposes, using
\eqref{showerLength} to 
model shower length of an electromagnetic shower in ice is thus sufficiently accurate. 

Based on these considerations of shower length, we model the TR signal coming from 
electromagnetic showers to scale with shower energy as $E_\mathrm{sh}^{2/3}$ to keep the product of 
shower length and number of particles 
at maximum linearly proportional to $E_\mathrm{sh}$, as discussed above. 
The largest value of $E_\omega d$ observed for a vertical EeV electron-induced shower
in our previous simulations \cite{Motloch:2015wca} was $1.5\, \mathrm{mV/MHz}$. We will
thus take this value as a maximal attainable strength of a TR signal from an EeV shower and
extrapolate this value to higher energies as
\be
\label{ElmagScaling}
	\langle E_\omega d\rangle_\mathrm{em}^\mathrm{max} = 1.5  
	\(\frac{E_\mathrm{sh}}{\mathrm{EeV}}\)^{2/3} \, \frac{\mathrm{mV}}{\mathrm{MHz}}.
\ee

On the other hand, the length of the hadronic showers are
mostly unaffected by the LPM effect because the hadronic interactions of the shower particles rapidly
degrade the energy before high energy electrons and photons are produced. The
shower length rises by only about 20\% as the primary shower
energy rises from 10 PeV to 10 EeV \cite{AlvarezMuniz:1998px}. We then model the shower
length of hadronic showers as constant
\be 
\label{HadronicLength}
	L_\mathrm{sh,had} = 5\,\mathrm{m} .
\ee 
Following the same reasoning as before we assume that the TR signal
  scales linearly with the shower energy.
Again, our final conclusions are mostly insensitive to the particular value of $L_\mathrm{sh,had}$. 
For low energies an electron-induced shower is assumed to be a good proxy for
a hadronic shower of the same energy as most of the energy of both electromagnetic and hadronic 
showers is ultimately dissipated in electromagnetic processes and the number
of particles at maximum are similar~\cite{AlvarezMuniz:1998px}. 
Using the maximal achievable value of the distance-scaled TR electric
field $6.4 \cdot 10^{-4}\, \mathrm{mV}/\mathrm{MHz}$ for a 100 TeV
electromagnetic shower found in \cite{Motloch:2015wca} and extrapolating it linearly with
shower energy we obtain
\be
\label{HadronicScaling}
	\langle E_\omega d\rangle_\mathrm{had}^\mathrm{max} = 6.4 
	\(\frac{E_\mathrm{sh}}{\mathrm{EeV}}\) \, \frac{\mathrm{mV}}{\mathrm{MHz}}.
\ee
Note that we can not extrapolate the result from the EeV simulations here, because those are already
affected by the LPM effect. The result in \eqref{HadronicScaling} represents the
highest achievable signal from a hadronic shower of energy below about
100 EeV at which the LPM effect is not expected to affect the shower length and it applies for
showers which cross the ice-air boundary at around the shower maximum. Showers which cross
away from the shower maximum would have their signal diminished.

All three neutrino flavors can interact through NC and CC interactions. 
In the former, the breakup of the nucleon induces a hadronic shower that
typically carries 20\% of the incoming neutrino energy. With the signal strength
\eqref{HadronicScaling} we then need at least a neutrino of energy $E_\nu \simeq 30$ EeV 
to produce the emission seen in ANITA. We stress out that this lower bound on the
energy -- and those which follow -- was obtained by extrapolating simulation results of lower energy showers
to showers initiated by $\sim$20-30 EeV particles, which are computationally expensive to
simulate. This bound thus serves only as a rough estimate of
neutrino energies necessary to produce the detected signal.

The outcome of the CC interactions
depends on the neutrino flavor. Charged current interactions of the
electron neutrino produce two showers, one electromagnetic and one hadronic carrying about
80\% respectively 20\% of the total energy \cite{Connolly:2011vc}. When this neutrino
interacts very close to the surface such that both showers cross the interface with
significant numbers of particles, the signal strength can be up to
\ba
\label{FullScaling}
	\lefteqn{\langle E_\omega d\rangle^\mathrm{max}} \\
&&	= \(6.4 
	\(\frac{0.2 E_\nu}{\mathrm{EeV}}\) + 1.5 
	\(\frac{0.8 E_\nu}{\mathrm{EeV}}\)^{2/3}\) \, \mathrm{\frac{mV}{MHz}}. \nonumber
\ea
Thanks to the electromagnetic contribution, the minimal neutrino energy now decreases to
$E_\nu \simeq 20$ EeV. 

Charged interactions of tau and muon neutrinos produce a hadronic shower carrying
again about 20\% of the neutrino energy and an outgoing
lepton. The muon lepton gradually loses its energy through stochastic processes that induce many low
energy showers over a very long path length \cite{Stanev:1989bc}. They can be discarded in this
work, because the number of particles in these showers, $\simeq 3 \cdot
10^{-5}(E_\mu / \mathrm{1\,GeV})$ \cite{Stanev:1989bc}, is very small compared to those
necessary to produce a sizable radio signal. Before decaying, the muon typically loses most of its
energy through ionization and stochastic processes and it 
does not produce a significant electromagnetic
shower on its decay. We thus assume only the hadronic shower from CC interactions of muon
neutrino can be detected.

The behavior of the tau lepton produced at the neutrino interaction vertex can in principle
be different as tau lepton has a shorter lifetime than muon. However, a 10 EeV tau
lepton with a decay length of $\sim$500 km loses half of its energy in about
5 km of ice through stochastic processes (using the analytic model III of \cite{Dutta:2005yt}). 
For those energies only about 1\% of the tau decays would induce a large
  enough shower to produce a TR signal comparable to the observed signal.
The contribution of tau neutrino CC interactions are thus subdominant and will be
neglected.
Only the hadronic shower from CC interactions of tau neutrino will be
considered. In both $\nu_\mu$ and $\nu_\tau$ CC
interactions, the minimal neutrino energy necessary to produce detectable signal
is therefore $E_\nu\simeq 30$ EeV.

We conclude that transition radiation from a neutrino-induced shower can in principle
explain the main characteristics of the observed signal for 
neutrino energies $E_\nu \sim 20 - 30$ EeV or higher.
It is known that for
these extreme energies, neutrinos are absorbed in the Earth unless their zenith angles are
very close to horizontal, $\theta_\mathrm{sh} \approx 90^\circ$ \cite{Gandhi:1995tf}. 
However, because of the
wide angular emission of transition radiation, Earth-skimming neutrinos
can induce a signal which is seen at a large angle below the horizon.

The fact that the detectable neutrinos by means of TR have to be Earth-skimming 
would help in diminishing the uncertainty on the direction of the incoming neutrino as determined from
Fig.~\ref{fig:polarization}, leading to a pointing resolution of a few degrees.  On the other hand, there is
a degeneracy between the neutrino energy and the stage of the shower development at which
the shower crosses the boundary \cite{Motloch:2015wca}, leading to a poor energy resolution.

\section{Flux constraints}

Having determined that transition radiation can be responsible for the
observed ANITA event, we evaluate whether the neutrino flux necessary to
explain the event in terms of TR satisfies the current experimental limits. We do so
in the current section by estimating an upper limit on the
exposure achieved by the ANITA balloon for observation of TR events and comparing it with
the exposures from other experiments. 

The three neutrino flavors and their NC and CC
interactions represent six channels which can lead to an experimental
signal. They are independent and the relevant calculations are very similar (or identical)
for all six cases. We thus first present in greater detail the calculation of the exposure for
the NC interaction of any of the flavors and later comment on how the
calculation has been altered to accommodate the CC interactions.

\subsection{Neutral current interaction}

The number of NC interaction events of any given neutrino flavor
detected by the experiment is a Poisson random variable. Its expectation value is
expressed by an integral over the neutrino energy
\be
\label{ExpectedEvents}
	\langle N \rangle = T \int dE_\nu\, F(E_\nu) A(E_\nu) ,
\ee
where $T$ is the duration of the measurement, $F(E_\nu)$ is the energy-dependent flux of neutrinos
and $A(E_\nu)$ is the aperture of the experiment.

The aperture of the ANITA experiment for TR detection can be written as
\be
	A(E_\nu) = \int_{S_\mathrm{eff}(E_\nu)} dS \int_{\mathrm{up-going}} d\Omega\,\,
	p_\mathrm{det} \cos \theta_\mathrm{sh}.
\ee
Here we integrate the probability $p_\mathrm{det}$ that the experiment detects a
particular neutrino described by its direction and point at which its trajectory leaves
the Earth\footnote{Or possibly its extension
if the neutrino interacts before the crossing point.}. The angular
integral goes over half of the sky to capture only 
the up-going neutrinos. The outer
integral goes over the surface of the Earth which is visible from the balloon and that is
close enough such that the detection is in principle possible (see further). The
additional factor of $\cos \theta_\mathrm{sh}$, where $\theta_\mathrm{sh}$ is the angle between the surface
normal and the neutrino direction, represents the attenuation of the neutrino flux 
due to the projection of the surface as seen by the neutrinos.

The probability $p_\mathrm{det}$ has three components
\be
	p_\mathrm{det} = p_t \cdot p_i \cdot p_p ,
\ee 
each assuming that the previous process occurred successfully:
\begin{itemize}
	\item Probability $p_t$ for the neutrino to cross through the Earth.
	\item Probability $p_i$ for it to interact close enough to the surface so that the induced 
        shower crosses the interface and produces
	sufficiently strong transition radiation at least in some directions on the sky.
	\item Probability $p_p$ to have the balloon placed at a position where the transition
	radiation is sizeable.
\end{itemize}

The probability for a given neutrino to go through the Earth is related to the total neutrino
interaction cross section on nucleons $\sigma_T$ \cite{Connolly:2011vc}, and the number
density of nucleons given by the density of Earth at distance $r$ from the center
$\rho(r)$ divided by the nucleon mass $M_n$ \cite{Gandhi:1995tf}. We model the Earth as
having an additional 2 km layer of ice with density $\rho_\mathrm{ice} =
0.924\,\mathrm{g/cm^3}$ to account for Antarctic ice; the Earth radius $R$ 
includes height of this ice cap. 
The probability can be then calculated as
\be
	p_t = \exp\(-\frac{\sigma_T}{M_n}\int_{-R \cos \theta_\mathrm{sh}}^{R
	\cos\theta_\mathrm{sh}}
	\rho\(\sqrt{R^2 \sin^2 \theta_\mathrm{sh} + l^2}\) dl\),
\ee
where the integral in $l$ goes along the neutrino trajectory within the Earth. Notice that the
probability depends implicitly on the neutrino energy through $\sigma_T$ and explicitly on
the neutrino inclination $\theta_\mathrm{sh}$. For neutrinos of EeV energies $p_t$
is practically zero except for a small solid angle region a few
degrees away from $\theta_\mathrm{sh} = 90^\circ$.

For transition radiation to be sizeable, we need the shower to leave the ice at a point
where the shower contains a large number of particles. Because of this, we model the
probability for a sufficiently close interaction as
\be
\label{PInteract}
	p_i = \frac{L_\mathrm{sh,had}\, \sigma_{NC} \,
	\rho_\mathrm{ice}}{M_n},
\ee
which is the probability for the neutrino to interact through a NC
interaction in a length comparable to the 
longitudinal dimension $L_\mathrm{sh,had}$ of the hadronic shower created by 
this interaction~\eqref{HadronicLength}. Here $\sigma_{NC}$ is the neutrino cross
section for NC interaction on nucleons.

From our previous discussion we know that for a 1 EeV neutrino the transition radiation is
not strong enough to produce a signal above the ANITA threshold which can
  be estimated to be 
$E_\mathrm{thr} \approx 150\, \mathrm{\mu V/m}$. This magnitude can be
    obtained considering the lowest 
    pulse amplitude of the four published events~\cite{Gorham:2016zah} and dividing it by a
  factor of order two. This is close to an estimate obtained for a horn antenna of 
  central frequency 600 MHz if we conservatively consider a detection threshold
  of 2.3 times the lowest RMS noise
  voltage of 10~$\mu$V quoted for ANITA~\cite{Gorham:2008dv}.
Assuming linear scaling of the radiated electric-field with energy 
and taking into account that only about 20\% of the neutrino energy goes
into the shower, the maximal distance at which TR emission of any given shower can be detected is
\be	
	D \approx 8.5 \(\frac{E_\nu}{\mathrm{EeV}}\) \, \mathrm{km} .
\ee
This sets a cutoff for the lowest theoretically detectable neutrino energy at
$\sim$ 4 EeV, and this value would only apply for vertical emission.

Finally, assuming the neutrino crosses through the Earth and interacts sufficiently close to
the surface at a point which is at a distance smaller than $D(E_\nu)$ from the balloon, we will assume
that the ensuing emission is detectable from any point in the sky, $p_p = 1$. This is not
a realistic assumption as for very inclined showers the emission is far from being isotropic
\cite{Motloch:2015wca}. We expect this overestimates the calculated aperture by about a
factor of 4-5, however this assumption vastly simplifies the calculations and is
sufficient to lead to physically relevant conclusions. Additionally, the ANITA
experiment can not detect signals from too close to the payload as its antennas are
observing the horizon. Neglecting this effect as we do increases the obtained aperture and thus
leads to an even more conservative upper bound.

With all these assumptions, the expression for the aperture decouples into
\be
\label{simplifiedAperture}
	A = \(\int_{S_\mathrm{eff}(E_\nu)} dS\) \(\int_\mathrm{up-going}p_i  p_t
             \cos \theta_\mathrm{sh} \, d\Omega\) 
\ee
as $p_p$ then depends only on the point where neutrino trajectory leaves the Earth
and $p_t, p_i$ depend only on the neutrino direction relative to the ground.
The first integral evaluates to the area which is visible from the balloon and which is
closer than $D(E_\nu)$. A straightforward calculation leads to
\be
\label{Seff}
	\int_{S_\mathrm{eff}(E_\nu)} dS =  
	 \begin{cases}
      0, & E_\nu < 4\, \mathrm{EeV} \\
		\frac{2 \pi R^2 h}{R + h} , & E_\nu >
		77 \,\mathrm{EeV} \\
      \frac{\pi R}{R+h}\left[D(E_\nu)^2 - h^2\right], & \text{otherwise}
    \end{cases} .
\ee
At the lowest energies the effective detection area is zero because the showers do not
have enough energy to produce a detectable electric field. At the highest energies the whole patch of the Earth of
area $ 1.35\cdot 10^6\, \mathrm{km^2}$ which is observable from the balloon is
sufficiently close to ascertain a detection. Intermediate energies interpolate between these two
limits. 

The second integral in \eqref{simplifiedAperture} can be rewritten as
\ba
	\Omega_\mathrm{eff} &=& \int_\mathrm{up-going} p_i  p_t  \cos
	\theta_\mathrm{sh} \, d\Omega\\
		&=& 2\pi \int_0^{1} p_i  p_t  \cos \theta_\mathrm{sh}\,
		d(\cos\theta_\mathrm{sh}) .
\ea
Using the known neutrino cross sections and density profile, we can evaluate this integral
numerically.

\begin{figure}
\center
\includegraphics[width = 0.48 \textwidth]{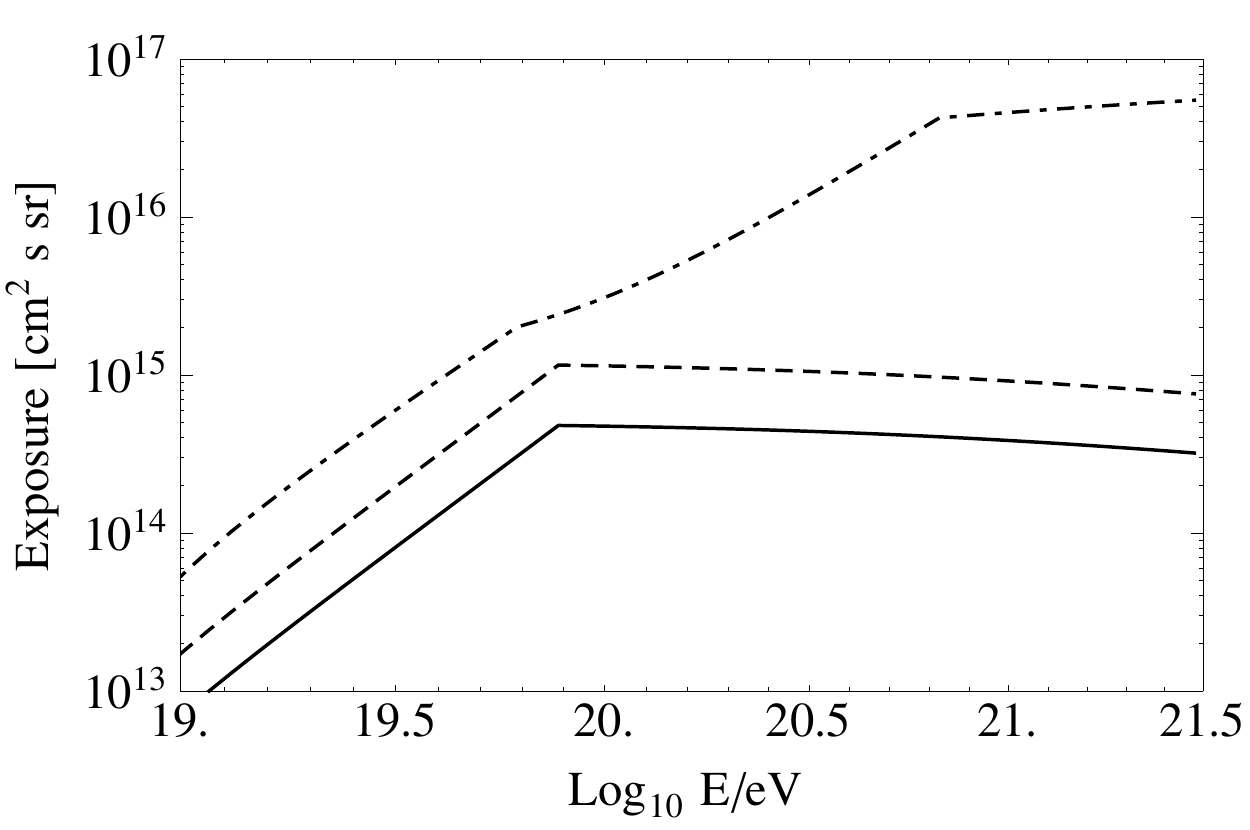}
\caption{Contributions to the conservative upper limit on the exposure for TR
detection from neutral current interaction (solid) and charged current interaction of
$\nu_\tau, \nu_\mu$ (dashed) and $\nu_e$ (dot dashed); each curve
represents a contribution of a single flavor.}
\label{fig:exposureSingle}
\end{figure}

With known $S_\mathrm{eff}$ and $\Omega_\mathrm{eff}$ we can calculate the exposure $T_a
A$
from the total live time $T_a = 45.8$ days of the two ANITA flights
\cite{Gorham:2010kv,Gorham:2008yk}. The result is plotted in
Fig.~\ref{fig:exposureSingle}.

\subsection{Charged current interactions}

As discussed previously, the hadronic showers from CC
interactions of tau and muon neutrinos are expected to produce significant contributions
to the number of detected events. This allows us to directly use the results of the
previous section, after replacing the NC interaction cross section in
Eq.~\eqref{PInteract} with the CC one.

For the electron neutrino, two showers are produced as discussed in the first section of the paper
and this requires a slightly more involved approach. When the neutrino interacts
very close to the surface, both these showers cross the surface and their emission adds.
Following our strategy of overestimating the aperture we assume they add
coherently,
leading to the TR signal in Eq.~\eqref{FullScaling}. For this to happen, the neutrino has to interact
in a length roughly $L_\mathrm{sh,had}$. In addition, there is a possibility that the neutrino
interacts in such a way that the electromagnetic shower elongated by the LPM
effect reaches the boundary, but the shorter hadronic shower does not. The length corresponding to
this situation is approximately $L_\mathrm{sh,em} - L_\mathrm{sh,had}$ and the emitted TR in
this case is just the second term in \eqref{FullScaling}. Using again the cross section for
CC interaction in Eq.~\eqref{PInteract}, we can evaluate the corresponding
exposure for CC interaction of $\nu_e$ in a way analogous to the previous section.

The energy dependence of the exposure from CC interactions is 
shown in Fig.~\ref{fig:exposureSingle}. The CC interactions of $\nu_\mu,
\nu_\tau$ lead to a higher exposure than the NC interactions due to
higher cross section for CC interactions. Charged current interactions of $\nu_e$ are largest everywhere
due to the additional electromagnetic contribution. This contribution dominates above
$10^{20}$ eV because of the increased length of the electromagnetic shower.

\subsection{Total exposure}

\begin{figure}
\center
\includegraphics[width = 0.48 \textwidth]{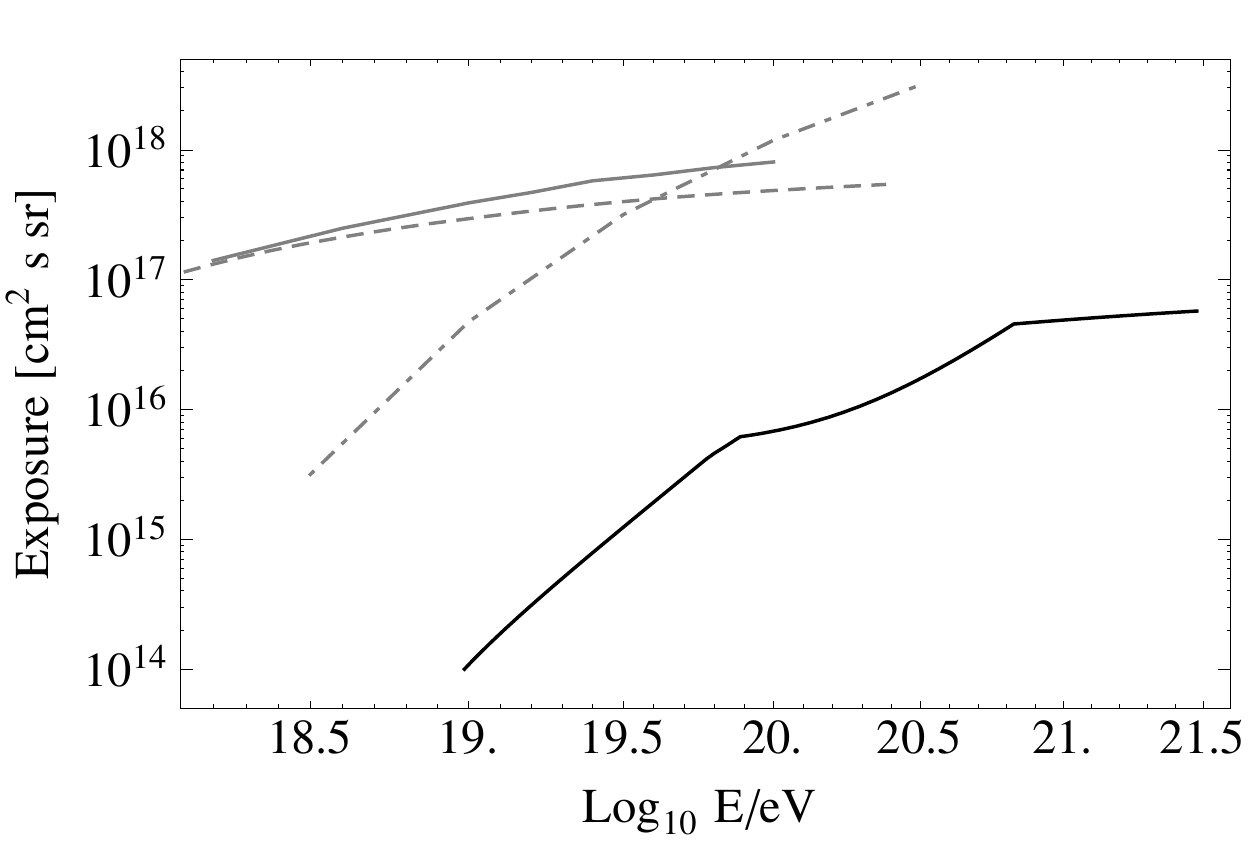}
\caption{Conservative upper limit on the exposure of the two combined ANITA flights for neutrino detection using TR (black),
compared with exposures of other experiments (gray): standard analysis of ANITA using
Askaryan radiation \cite{Mottram:2012jsa} (dot-dashed), Pierre Auger Observatory
\cite{Aab:2015kma} (dashed), IceCube \cite{Aartsen:2015zva, Aartsen:2013dsm} (solid). } 
\label{fig:exposure}
\end{figure}

Adding the contributions plotted in Fig.~\ref{fig:exposureSingle} for all six combinations of flavor and 
interaction channels, an upper limit for the total exposure of the two first ANITA flights for neutrino
detection with the TR is plotted in Fig.~\ref{fig:exposure}.

In the same plot we also display the exposures of other experiments searching for cosmogenic
neutrinos: the analysis of ANITA when trying to detect Askaryan radiation from neutrino-induced
showers in ice \cite{Mottram:2012jsa}, assuming the same live time $T_a$;
approximately nine years of data collected at the Pierre Auger Observatory \cite{Aab:2015kma};
and preliminary results of six years of IceCube \cite{Aartsen:2015zva}. For IceCube we used the neutrino
effective area from \cite{Aartsen:2013dsm} with data taken in several periods of time with 
various string configurations \cite{Aartsen:2015zva, Aartsen:2013dsm}.

\section{Discussion}
\label{sec:discuss}

In this paper we evaluated the possibility that the signal of an unknown origin detected by
the ANITA collaboration is a transition radiation from a shower induced by a high-energy  
neutrino. While the properties of the observed signal can in principle be explained with a 
transition radiation origin for the signal, the flux necessary for successful explanation 
is in tension with the
current best limits from the Pierre Auger observatory, IceCube and ANITA. In a
conservative analysis we found that the exposure for the TR
signal is at least two orders of magnitude smaller than the exposure collected by 
ANITA at the energies necessary to explain the event in terms of transition radiation;
this exposure is also smaller than extrapolations of the Pierre Auger and IceCube
exposures to these energies. The probability that ANITA detected transition radiation induced
by a cosmogenic neutrino shower and none of the other experiments detected any neutrino
event is thus negligible. 

Our analysis of the aperture of the experiment for the TR signal was optimistic by
assuming fully isotropic emission. Moreover,  due to its
experimental design ANITA cannot observe signals produced right below the
payload, but this was ignored in the exposure estimate.
In reality, the exposure for a TR detection by the
ANITA balloon is significantly smaller, further decreasing the probability of a TR explanation of
the ANITA event. 

From a comparison of the exposures of the TR signal with the standard ANITA analysis we
also see that TR is not a viable alternative for this particular geometry and presents
less than $\sim$1\% correction to the exposure. This is caused by the fact that the traditional Askaryan
emission searched for by ANITA is stronger than transition radiation. Moreover the 
detection technique exploiting the Askaryan radiation benefits from the fact that a detectable neutrino
interaction can occur at distances of $\sim$ km from the ice-air interface 
and not only close to the surface,
further disfavoring the transition radiation. This however does not mean that for a different
geometry --- such as a ground array --- TR does not pose a viable method of detection and
further studies in this direction are necessary.

\acknowledgements{
This work was supported by U.S.~Dept.\ of Energy
contract DE-FG02-13ER41958. 
J.A-M and E.Z thank Ministerio de Econom\'\i a (FPA 2015-70420-C2-1-R), 
Consolider-Ingenio 2010 CPAN Programme (CSD2007-00042),  
Xunta de Galicia (GRC2013-024), Feder Funds,  
Marie Curie-IRSES/ EPLANET (European Particle physics Latin
American NETwork), $7^{\rm th}$ Framework Program (PIRSES-
2009-GA-246806) and RENATA Red Nacional Tem\'atica de Astropart\'\i culas 
(FPA2015-68783-REDT). This work was supported in part by the Kavli Institute for
Cosmological Physics at the University of Chicago through grant NSF PHY-1125897 and an
endowment from the Kavli Foundation and its founder Fred Kavli.
}

\bibliography{anita}

\end{document}